\documentclass[twocolumn,a4paper,superscriptaddress,showpacs,showkeys]{revtex4-1}
\usepackage{graphicx}
\usepackage{subfigure}
\usepackage{latexsym}   
\usepackage{amsmath,amssymb,amsthm}

\usepackage{listings}

\usepackage{fancyvrb}  

\begin{document}

\title{Microcanonical thermostatistics analysis without histograms: cumulative distribution and Bayesian approaches}


\author{\firstname{Nelson} A. \surname{Alves}} %
\affiliation{Departamento de F\'{\i}sica, FFCLRP, Universidade de S\~ao Paulo, Avenida Bandeirantes, 3900, 14040-901, Ribeir\~ao Preto, SP, Brazil.}
\author{\firstname{Lucas} D. \surname{Morero}} %
\affiliation{Departamento de F\'{\i}sica, FFCLRP, Universidade de S\~ao Paulo, Avenida Bandeirantes, 3900, 14040-901, Ribeir\~ao Preto, SP, Brazil.}
\author{\firstname{Leandro} G. \surname{Rizzi}} %
\affiliation{Departamento de F\'{\i}sica, FFCLRP, Universidade de S\~ao Paulo, Avenida Bandeirantes, 3900, 14040-901, Ribeir\~ao Preto, SP, Brazil.}
\affiliation{School of Chemistry, University of Leeds, LS2 9JT, Leeds, UK.}


\date{\today}


\begin{abstract}
\noindent
    Microcanonical thermostatistics analysis has become an important tool to reveal essential 
aspects of phase transitions in complex systems. 
    An efficient way to estimate the microcanonical inverse temperature $\beta(E)$ and the microcanonical 
entropy $S(E)$ is achieved with the statistical temperature weighted histogram analysis method (ST-WHAM).
    The strength of this method lies on its flexibility, as it can be used to analyse data produced by algorithms 
with generalised sampling weights.
    However, for any sampling weight, ST-WHAM requires the calculation of derivatives of energy histograms 
$H(E)$, which leads to non-trivial and tedious binning tasks for models with continuous energy spectrum 
such as those for biomolecular and colloidal systems.
     Here, we discuss two alternative methods that avoid the need for such energy binning to obtain continuous estimates 
for $H(E)$ in order to evaluate $\beta(E)$ by using ST-WHAM: 
(i) a series expansion to estimate probability densities from the empirical cumulative distribution function (CDF), and
(ii) a Bayesian approach to model this CDF. 
    Comparison with a simple linear regression method is also carried out.
    The performance of these approaches is evaluated considering coarse-grained protein models for folding and peptide 
aggregation.
\end{abstract}


\maketitle

\noindent

\section{Introduction}
   Fundamental aspects of phase transitions in complex systems can be revealed by the analysis of its microcanonical 
thermostatistics~\cite{gross-book,gross2}, which is characterised by the well known entropy $S(E)=k_B \ln \Omega(E)$, 
where $\Omega(E)$ denotes the density of states of a system with energy $E$.
   In particular, the analysis of inflection points of the microcanonical inverse temperature 
$\beta(E)=dS(E)/dE$ plays an important role in the identification of stable, unstable and metastable 
regions in the phase diagram~\cite{frigori2010JCS,schnabelPRE2011,rochaPRE2014}, providing alternative insights to 
the usual canonical analysis.
   Also, free-energy profiles can be obtained from the caloric curves $\beta$ vs $E$, from where one can easily evaluate 
the values of barrier heights and latent heats.
   In this way, the microcanonical thermostatistics analysis has been incorporated in many studies in the literature, 
{\it e.g.}~Refs.~\cite{frigoriJCP2013,jankePRL2006,jankeJCP2008,moddel2010pccp,bereau2010jacs,church2012jcp,straubPRL2014} 
to name a few.

   Nevertheless, any analysis must rely on data obtained from efficient exploration of the configurational space.
   It is well known that numerical simulations performed with Monte Carlo (MC) and molecular dynamics (MD) methods 
pose limitations to the achievement of reliable data sampling \cite{okamoto-60}.
   Such limitations are related to the critical slowing down \cite{landau-book}, which is observed in studies of continuous 
phase transitions, and to the entrapment in local minima, in the case of systems with rugged 
energy landscapes. 
   In both cases, the configurational space is poorly explored in a reasonable computational simulation
time, which may produce biased physical averages.								
   To overcome the trapping problem, it has been suggested that configurations must be sampled using algorithms
based on generalised ensembles, where the updates are performed with non-Boltzmann statistical weights $\omega(E)$.
   For instance, the multicanonical algorithm (MUCA) \cite{berg2000,berg2003}, the extended Gaussian ensemble (EGE) 
\cite{johal,neuhaus06,frigoriEurP2010}, Tsallis statistical weight \cite{straub-tsallis,hansmannPRE1997}, and 
replica exchange method (REM) \cite{hukushima1996} either use a series of Boltzmann weights or 
any convenient generalised sampling weight \cite{straub-gREM}.

   MUCA simulations sample configurations with a weight $\omega_{mu}(E)$ in such a way that the energy distribution
is uniform, $ H_{mu}(E) \propto \Omega(E) \, \omega_{mu}(E) = constant $.
   Thus, a precise determination of $\omega_{mu}(E)$ is equivalent to obtain an estimate for the density of states $\Omega(E)$,
{\it i.e.} $\omega_{mu}(E) \propto 1/\Omega(E)$.
   The weights  $\omega_{mu}(E)={\rm exp}[-b(E)E+a(E)]$ follows from the parameterization of the entropy
 $S(E) = b(E) E - a(E)$,  where $a(E)$ and $b(E)$ are the so-called multicanonical parameters.
   The iterative procedure to obtain the MUCA parameters is described in detail in references \cite{berg2000,berg2003}, 
and can be read as,
\begin{equation}
a^{n}(E_{m-1}) = a^{n}(E_{m}) + [b^{n}(E_{m-1})-b^{n}(E_{m})]E_{m}, \label{eq:an} \\
\end{equation}
\vbox{\begin{align} 
b^{n}(E_{m}) = b^{n-1}(E_{m}) + [\ln H^{n-1}_{mu}(E_{m+1}) -\ln H^{n-1}_{mu}(E_{m})]/ \varepsilon,           
			 \label{eq:bn} 
\end{align} }
where $n$ stands for the $n$th multicanonical simulation.
    The recursion steps require the discretisation of the energy for continuous energy models. 
    Therefore, it is convenient to define $E_{m}=E_{0} + m \varepsilon$, where $\varepsilon$ is the binsize,
$m$ is an integer, and $E_{0}$ is a constant that defines a reference energy.
    All the energies $E$ in the interval $[E_{m},E_{m+1}[$ contribute to the  histogram $H_{mu}(E_{m})$. 

    Methods to improve sampling based on simulations at different temperatures have been proposed to either be conducted in parallel (REM) or as a random walk between different temperatures.
   In REM, $N_{\rm rep}$ non-interacting replicas of the system  are simultaneously simulated by the usual MC or MD 
algorithms, and from time to time, pairs of replicas at neighboring temperatures are exchanged with a transition 
probability.
    From the data produced by simulations  performed at a single temperature $T_1$ or at a set of temperatures
$T_{\alpha}$, with $\alpha=1,2, \cdots, N_{\rm rep}$, it is necessary to employ a reweighing scheme to evaluate 
physical averages at a given temperature $T$. 
    Reweighting techniques \cite{ferrenberg-rewei,wham,alves92}
use data from either a single histogram or multiple histograms obtained from MC or MD simulations.
  
    Recently, a simple method called statistical weighted histogram analysis method (ST-WHAM)~\cite{kimJCP-stwham} 
has been proposed as an iteration-free procedure to obtain an estimate for the microcanonical inverse temperature.
    In this method the usual WHAM equations~\cite{ferrenberg-rewei,wham} are converted into a weighted average of 
the individual densities of states obtained from simulations carried out with different sampling weights
$\omega(E)$.
    From energy histograms produced by multiple simulations, ST-WHAM yields a statistical temperature 
${\tilde T}(E) = 1/{\tilde \beta}(E)$, which is an estimate of the inverse microcanonical temperature 
$\beta(E)=d S(E)/d E$.
   Interestingly, there is a numerical procedure based on the multicanonical recursion relations 
(\ref{eq:an}) and (\ref{eq:bn}), which is called ST-WHAM-MUCA  \cite{rizziJCP2011}, that can be used to 
replace the direct integration in order to evaluate the entropy $S(E)$.
   Although both ST-WHAM and ST-WHAM-MUCA have the advantage of {\it a posteriori} discretisation of energies, their
naive implementations may lead to biased evaluations of physical quantities for continuous energy models
just like all the aforementioned algorithms.
	
   As described in Section II, the estimates ${\tilde \beta}(E)$ for inverse microcanonical temperature $\beta(E)$ 
depends on the derivatives of the energy histograms $H(E)$ (see Eq. (\ref{eq:st-wham2})).
   Here, we analyse how the estimates ${\tilde \beta}(E)$ are energy 
binning dependent and, in Section III, we present two alternative approaches that avoids the need for energy binning 
to evaluate the microcanonical caloric curve for continuous energy models: 
(i) a proposal by Berg and Harris~\cite{berg-accum}, which involves an empirical cumulative distribution (CDF) and 
uses discrete Fourier series; and
(ii) a Bayesian approach~\cite{bayes-caldwell} to model this CDF and assumes that 
the thermodynamic phase transitions are well described by the coexistence of two phases. 
    A comparative analysis between these approaches is made in order to characterise $\beta(E)$ 
for two systems that undergo first-order-like phase transitions: the folding transition of a 
coarse-grained protein model for Ubiquitin 
and the aggregation transition of two heteropolymers that follows a Fibonacci sequence.
    These examples allow us to describe the statistical and systematic errors involved in the numerical calculations of $H(E)$ 
and $\tilde{\beta}(E)$, which are presented in Section IV.
	Conclusions on this comparative analysis are presented in Section V.

\section{Statistical temperature weighted histogram method}

   The ST-WHAM \cite{kimJCP-stwham} yields a direct estimate of the inverse microcanonical temperature
 $  \beta(E)  = d\, {\rm ln}\, \Omega(E)/d E $                      
by considering the statistical inverse temperature
\begin{equation}
      {\tilde \beta}(E) =  \sum_{\alpha} f_{\alpha}^*(\beta_{\alpha}^H + \beta_{\alpha}^\omega) \, , \label{eq:st-wham1}
\end{equation}
where 
   $f_{\alpha}^* =H_{\alpha}/\sum_{\gamma}H_{\gamma},$ $\beta_{\alpha}^H = d\, {\rm ln} H_{\alpha}/{d E}$,
and 
   $\beta_{\alpha}^\omega= -d\, {\rm ln}\, \omega_{\alpha}/{d E}$.
    It is preferable to write Eq. (\ref{eq:st-wham1}) as
\begin{equation}
 {\tilde \beta}(E) =
     \frac{1}{\sum_{\gamma}H_{\gamma}(E)}
		          \sum_{\alpha} H_{\alpha}(E)\left( \frac{d\,{\rm ln}H_{\alpha}(E)}{d E} - 
              \frac{d\, {\rm ln}\omega_{\alpha}(E)}{d E}\right)     \, .              \label{eq:st-wham2}
\end{equation}							
    Note that  $\beta_{\alpha}^{\omega}=1/T_{\alpha}$ for simulations with the canonical weight.
    With the set of estimates ${\tilde \beta}(E_m)$, MUCA recurrence relations (\ref{eq:an}) and (\ref{eq:bn}) 
can be applied to obtain estimates ${\tilde S}(E_m)$ for the microcanonical entropy $S(E_m)$,
\begin{equation}
  {\tilde S}(E_m) = {\tilde \beta}(E_m) E_m - a(E_m) \, .                         \label{eq:st-wham-muca}
\end{equation}
   This ST-WHAM-MUCA algorithm is quite simple if one has ${\tilde \beta}(E_m)$.

\section{Numerical Evaluation of Derivatives}

\subsection{Linear regression}

    We can numerically evaluate the derivatives in Eq. (\ref{eq:st-wham2}) in a naive way, where the 
derivatives $d {\rm ln} H(E)/d E$ at energies $E_m$ follow from a linear regression around this point.
    For instance, we use a linear regression with $k=15$ points; selecting 
$k$ points means that the derivative at $E_m$ is calculated with the values of $H(E_{\ell})$, where 
$\ell = \mbox{$m-(k-1)/2$},  \mbox{$m+1-(k-1)/2$}, \cdots, m, \cdots, m+ \mbox{$(k-1)/2$}$.
    We chose a value for $k$ according to the energy binsize $\varepsilon$.
    Consequently, we calculate the derivatives in the energy range $\Delta E = (k-1) \varepsilon$.
    In this method, it is more convenient to directly calculate the derivative of ${\rm ln} H(E_m)$ than the derivative
of $H(E_m)$. 
    We calculate the linear regression with a subroutine easily adapted from the linear fit subroutine in \cite{recipes}.

\subsection{Cumulative distribution method}

    Another approach can be devised by considering an algorithm based on the cumulative distribution function 
(CDF)~\cite{berg-accum}.
    The advantage of such approach is that it avoids histogramming when describing probability densities $P(E)$, dismissing 
the need for any {\it ad hoc} energy discretisation.
    The method defines an estimator ${\tilde F}(E)$ for the CDF $F(E)$, where the function ${\tilde F}(E)$ is an 
empirical cumulative distribution function (ECDF) for the probability density $P(E)$.
    The algorithm sorts the energy time series of length {\ttfamily NDAT} in an ascending order 
$(E_1 < E_2 < \cdots < E_{\rm NDAT})$, so any outliers can be eliminated by constructing a restricted ECDF 
${\tilde F}_{ab}(E)$ in the range between two meaningful points $a$ and $b$ (in general one takes $a=E_1$ and $b=E_{\rm NDAT}$).
    The basic idea is to propose an approximating function $F_0(E)$ to describe ${\tilde F}_{ab}(E)$, from where 
the difference function is defined,
\begin{equation}
    R(E) = {\tilde F}_{ab}(E) - F_0(E) \, .                    \label{eq:re-def}
\end{equation}
This function can be expanded in Fourier series,
\begin{equation}
    R(E) = \sum_{m=1}^{\rm MMAX} d(m)\, {\rm sin} \left(\frac{m \pi(E-a)}{b-a}\right) \, , \label{eq:re}
\end{equation}
which gives the Fourier coefficients \cite{berg-accum},
\begin{equation}
    d(m) = \sqrt{\frac{2}{b-a}}\, \int_{a}^{b} R(E)\, {\rm sin} \left(\frac{m \pi(E-a)}{b-a}\right) dE \, . \label{eq:di}
\end{equation}
 Here we use the same criteria of~\cite{berg-accum} where the maximum number {\ttfamily MMAX} of coefficients $d(m)$ is 
obtained by imposing the two-sided Kolmogorov test~\cite{recipes}.

     Notably, Eq.~(\ref{eq:di}) provides all the information that one needs to obtain a parameter-free energy probability 
density $P(E)$.
     The reason is because Eq.~(\ref{eq:re-def}) yields a smooth estimate of the CDF given {\ttfamily MMAX} coefficients 
in the Fourier expansion,
\begin{equation}
    {\tilde F}_{ab}(E) = F_0(E) + R(E) \, ,
\end{equation}
 even if one assumes a linear {\it ansatz} for $F_0(E)$ (see \cite{berg-accum}).
     In turn, it becomes easy to obtain estimates for the probability density $P_{ab}(E)$ by differentiation, which consists 
in a smooth estimation of $H(E)$.
    
   Now, let us go back to the numerical differentiation in Eq. (\ref{eq:st-wham2}).
   To obtain parameter-free derivatives $d H(E)/d E$, we just need to 
differentiate ${\tilde F}_{ab}(E)$ again.
   This corresponds to the following numerical calculation,
\begin{equation}
  \frac{d^2 {\tilde F}_{ab}(E)}{d E^2} = 
	    - \sum_{m=1}^{\rm MMAX} d(m)\, \left(\frac{m \pi}{b-a}\right)^2 {\rm sin} 
			  \left(\frac{m \pi(E-a)}{b-a}\right) \, . \label{eq:derpd}
\end{equation}
    Unlike the linear regression method, where we calculate the derivative of ${\rm ln} H(E)$, here is more convenient to 
compute the derivative of $H(E)$ directly.
    We have included in Appendix the {\ttfamily FUNCTION DERPD(X)}, which can be used to estimate this derivative from the 
Fourier coefficients $d(m)$.
    One can easily incorporate this function in the program {\ttfamily CDF\_PD} subroutine provided by Berg and 
Harris~\cite{berg-accum}.

\subsection{Bayesian analysis}

    Next we present our approach to implement a Bayesian analysis.
    To this end,  input data (denoted by $\vec{y}$) is the empirical cumulative distribution  and not
the histograms that are dependent on the energy binsize $\varepsilon$.
    Thus, the aim is to describe the empirical cumulative distribution by a given model (called model M) and using
the Bayesian analysis \cite{bayes-caldwell}, extract values for its parameters (denoted by $\vec{\lambda}$)
from the {\it posterior} probability density function (PDF) $P(\vec{\lambda},\vec{\nu}\, |\, \vec{D}, M)$. 
    Here, $\vec{\nu}$ denotes possible parameters related to the experimental conditions, which is the
temperature in our numerical experiment, and $\vec{D}$ refers to a set of observed experimental points $\{E_i\}$.
    From the {\it posterior} PDF, we extract the desired probability distributions for each parameter,
\begin{equation}
 P(\lambda_i |\, \vec{D},M) = \int P(\vec{\lambda},\vec{\nu}\, |\, \vec{D},M)\, d \vec{\lambda}_{(j \ne i)}\, d\vec{\nu} ,
\end{equation}
which correspond to the marginalized distributions.
    These marginalized distributions are depicted with histograms in Fig. \ref{fig:6} for the Ubiquitin.

\begin{figure}[ht]
\begin{center}
\begin{minipage}[h]{0.45\textwidth}
\includegraphics[width=0.95\textwidth]{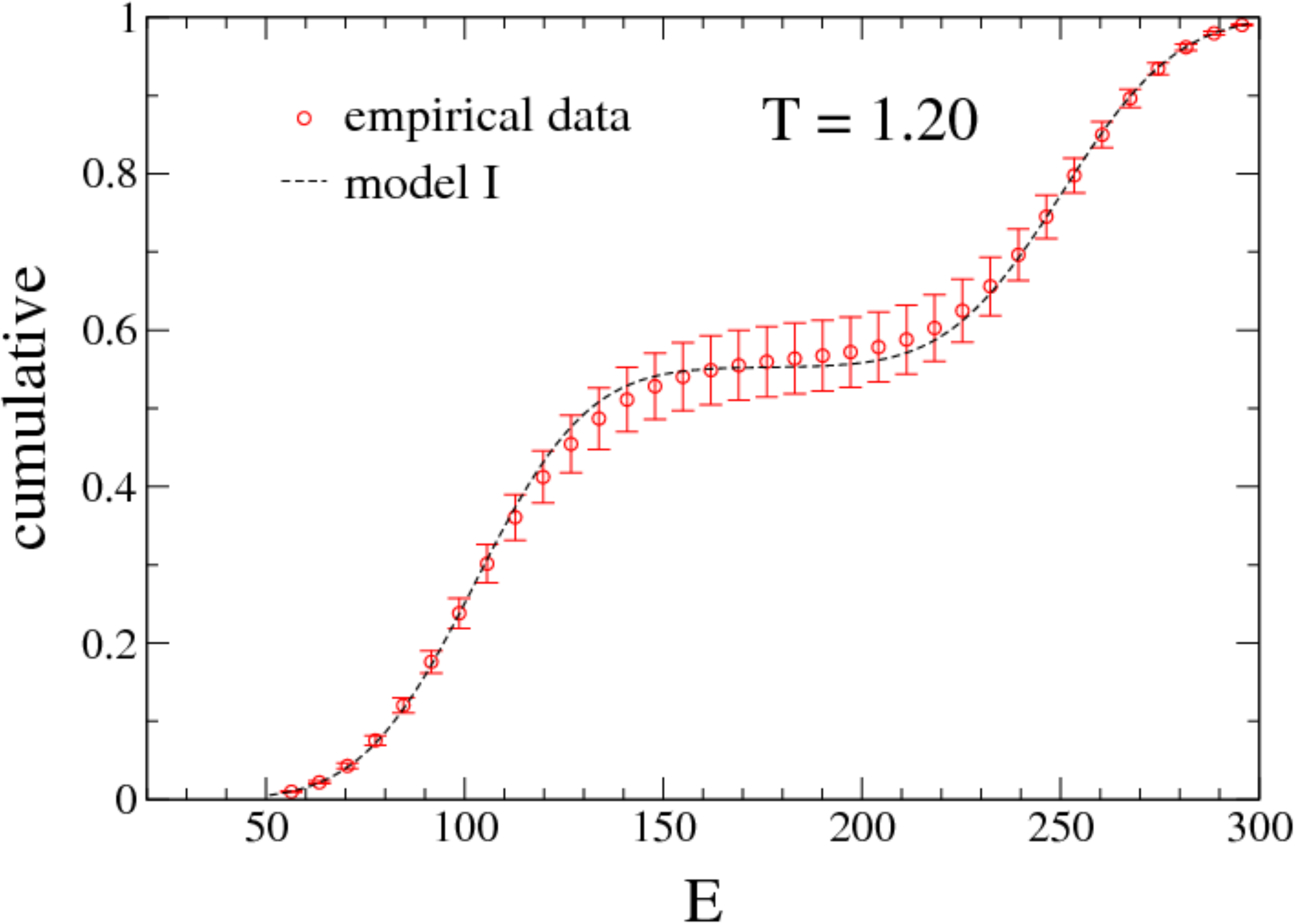}
\caption{Cumulative distribution data ($\circ$) of Ubiquitin with error bars calculated with 20 jackknife bins from
MD simulation at $T=1.20$. Dotted line corresponds to the Bayesian estimates of the cumulative distribution
assuming model I.}
\label{fig:4}
\end{minipage}%

\vspace{0.8cm}

\begin{minipage}[h]{0.45\textwidth}
\includegraphics[width=0.95\textwidth]{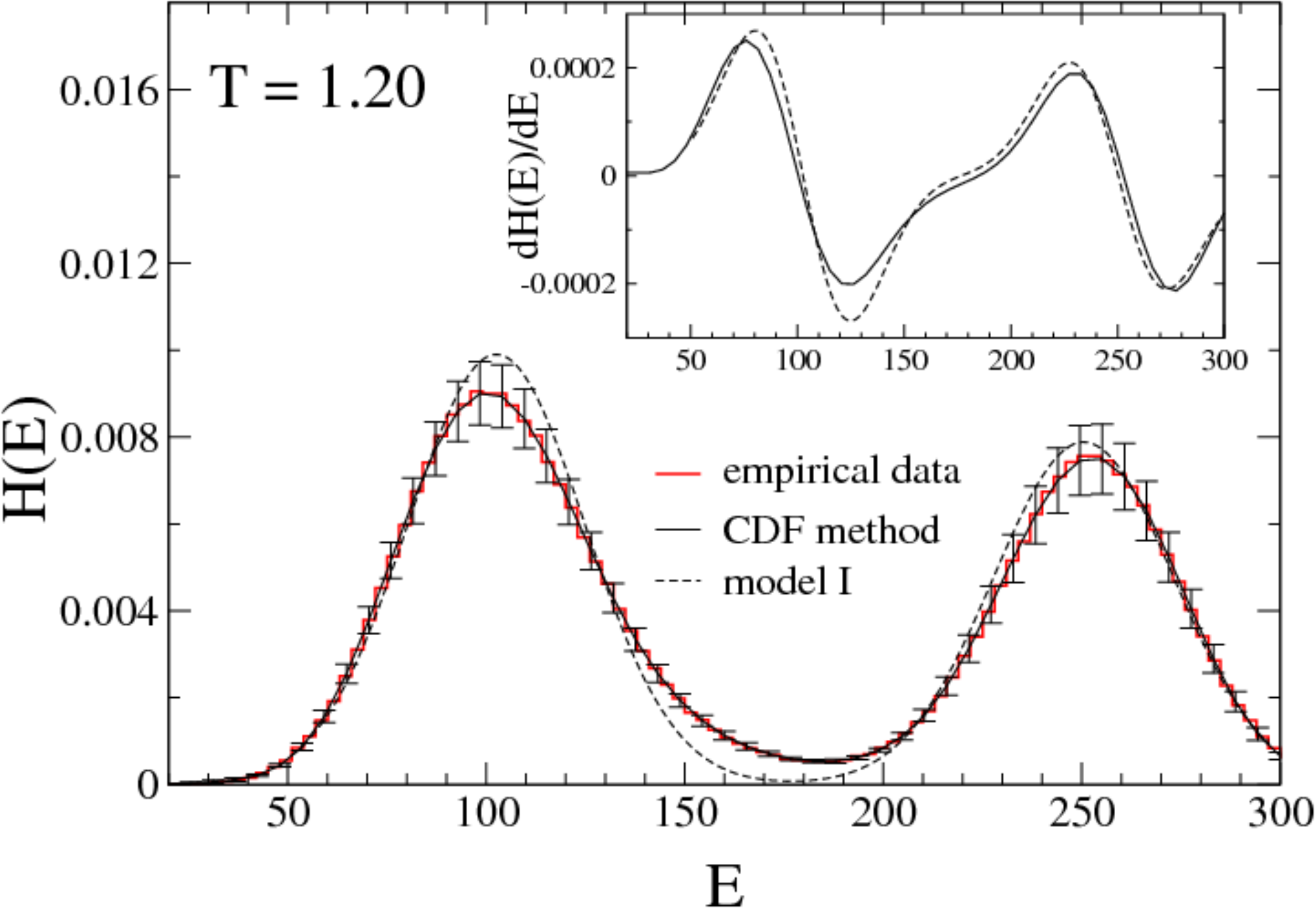}
\caption{Energy histograms $H(E)$ for Ubiquitin. Error bars are calculated
with 20 jackknife bins and are shown only for the CDF method. Inset: comparative
behaviour of $dH(E)/dE$ produced with the CDF method and Bayesian approach
assuming model I.}
\label{fig:5}
\end{minipage}%
\end{center}
\end{figure}

    For each protein, our {\it experimental} data $\vec{y}$ is produced by dividing
the observed energy range into 35 points $\{E_i\}$ and {\it experimental} errors are extracted from the jackknife 
procedure.
     We anticipate here that these {\it experimental} data are displayed in Figures \ref{fig:4} and \ref{fig:10} 
for the folding-unfolding transition of Ubiquitin and for the aggregation of two Fibonacci sequences, respectively.
     Within a Bayesian approach, the set of observations $\{y_i\}$ is just a set among possible experimental
outcomes $\vec{y}$ at the points $\{E_i\}$.
     Thus, given a fixed model M and a set of experimental values $\vec{y}(\{E_i\})$, parameters estimation
within Bayesian approach is obtained as follows  \cite{bayes-caldwell},
\begin{equation} 
 P(\vec{\lambda},\vec{\nu}\, |\, \vec{D},M) = 
        \frac{P(\{E_i\}=\vec{D}\, |\,\vec{\lambda},\vec{\nu},M)\, P_0(\vec{\lambda},\vec{\nu}\, |\, M)}
        {\int P(\{E_i\}=\vec{D}\, |\,\vec{\lambda},\vec{\nu},M)\, 
                  P_0(\vec{\lambda},\vec{\nu}\, |\, M)\, d \vec{\lambda}\, d\vec{\nu} },
\end{equation}
where $P_0(\vec{\lambda},\vec{\nu}\, |\, M)$ is the {\it prior} distribution for a fixed model M.

     Since our aim is to characterise $\tilde{\beta}(E)$ in a first-order phase transition region, 
we assume that the energy distribution $P(E)$ ({\it i.e.}~a continuous estimate for $H(E)$ in 
Eq.~(\ref{eq:st-wham2})) is well described by a double-peak function.
     This function characterises the coexistent phases at temperatures close enough to the transition 
temperature $T_c$, whose shape is a consequence of the free-energy barrier between the ordered and disordered phases.
     Therefore, the expected energy distribution $P(E)$ can be written as a normalized sum of two Gaussian distributions 
with peaks centered at energies $\mu_1$ and $\mu_2$, corresponding to the disordered and ordered phases, 
respectively \cite{Lee-1991}.
     If we assume the model M as the cumulative distribution function of $P(E)$, and recalling that for
a Gaussian distribution the CDF is the error function, we arrive to the following modelling,
\begin{equation}
  F(E) = 0.5 \left[ 1 +  a\, {\rm erf}\left(\frac{E-\mu_1}{s_1 \sqrt{2}}\right) 
                   + (1-a)\, {\rm erf}\left(\frac{E-\mu_2}{s_2 \sqrt{2}}\right) \right].   \label{eq:fe}
\end{equation}
    This is a 5-parameter model, which we call model I in our analysis, with
$ \vec{\lambda} = (\mu_1, s_1, \mu_2, s_2, a)$.

    Now, to obtain the {\it posterior} probability of the parameter set $\vec{\lambda}$,
given data set $\vec{D}$, from the {\it likelihood} $P_{i+1}(\vec{D}\, |\, \vec{\lambda})$ times the
initial {\it prior} probabilities $P_i(\vec{\lambda})$ for our model I, we use a Markov chain Monte
Carlo. 
    To this end, we consider the Metropolis algorithm with the proposal function
\begin{equation}
  P(\vec{D}\, |\, \vec{\lambda}) = \prod_{i=1}^{35} \, \frac{1}{\sigma_i \sqrt{2\pi}}\,
          e^{-\left[y_i - F(E_i)\right]^2/2 \sigma_i^2} \, ,
\end{equation}
where $\sigma_i$ are the errors in our computational experiment.
    The values $F(E_i)$ are calculated with the accepted values for the parameters $\vec{\lambda}$
of model I.

\section{Numerical Comparisons}

    In the following, we present results where we evaluate the performance of the different approaches in 
obtaining continuous estimates for $H(E)$ and, consequently, the estimate $\tilde{\beta}(E)$ for microcanonical 
inverse temperature.
    The comparative analysis is made with data obtained from both MC and MD simulations, where we employ two 
simplified off-lattice protein models, {\it i.e.} all atoms in the polypeptide chain are replaced by beads located 
at the $C_{\alpha}$-atom position, and present a continuous energy spectrum.

\begin{figure}[t]
\begin{center}
\begin{minipage}[h]{0.45\textwidth}
\subfigure{\includegraphics[width=0.95\textwidth]{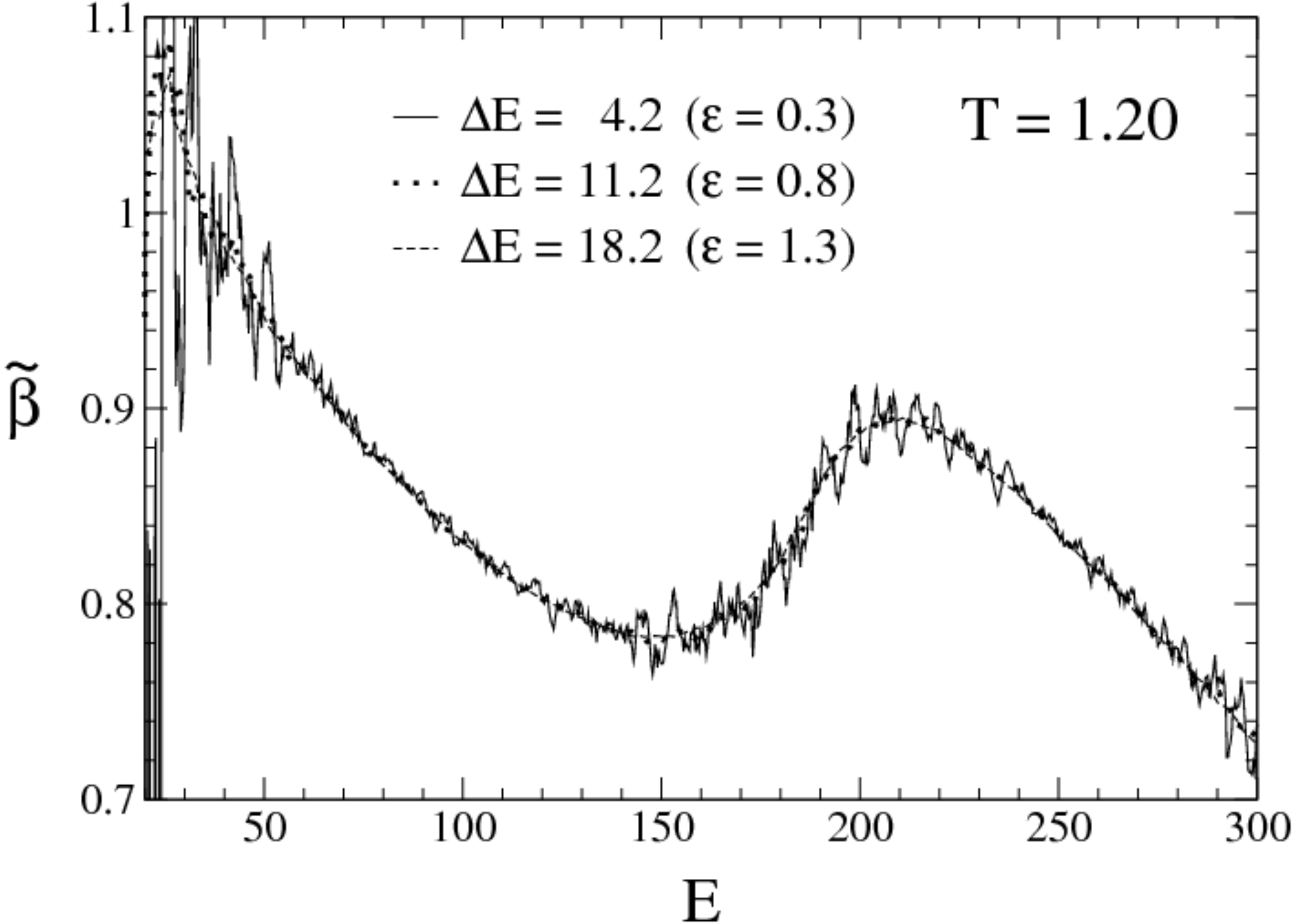}}
\caption{Statistical temperature estimates ${\tilde \beta}(E)$ of Ubiquitin using data 
obtained from MD simulations at $T=1.20$. Curves show the dependence of ${\tilde \beta}(E)$ on energy discretisations 
$\varepsilon$ ($\varepsilon= 0.3, 0.8$ and 1.3) for the linear regression method with $k=15$ points.}
\label{fig:1}
\end{minipage}%

\vspace{0.8cm}

\begin{minipage}[h]{0.45\textwidth}
\subfigure{\includegraphics[width=0.95\textwidth]{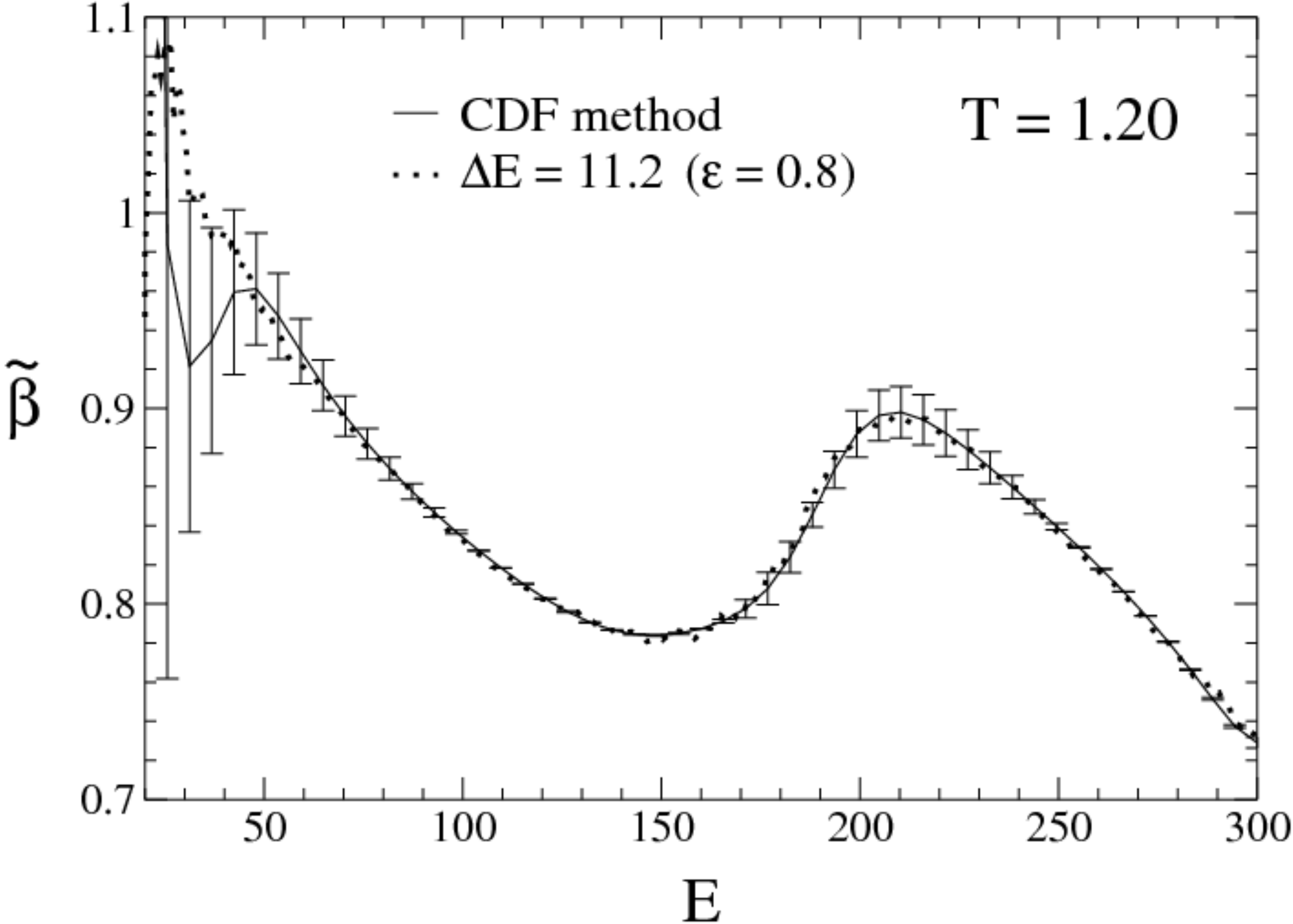}}
\caption{Comparison between estimates ${\tilde \beta}(E)$ of Ubiquitin with the linear regression 
($\varepsilon= 0.8$) and CDF method for data obtained at $T=1.20$. Statistical errors for the CDF method were calculated 
with 20 jackknife bins.}
\label{fig:2}
\end{minipage}
\end{center}
\end{figure}

\begin{figure}[ht]
\begin{center}
\includegraphics[width=0.45\textwidth]{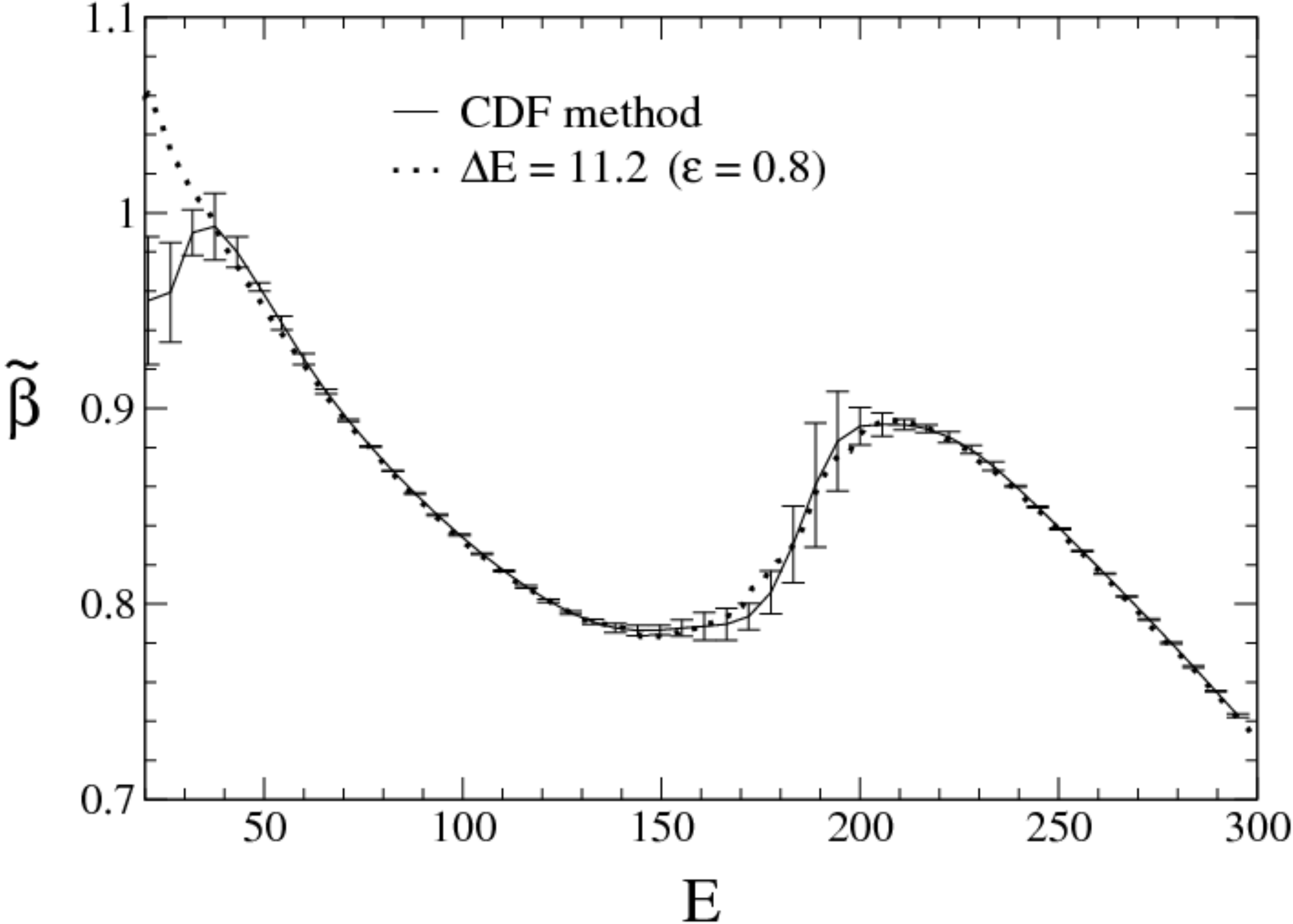}
\end{center}
\caption{Estimates ${\tilde \beta}(E)$ for Ubiquitin obtained with linear regression 
($\varepsilon= 0.8$) and CDF method with data generated from independent MD simulations 
at $T = \{ 1.14, 1.16, 1.18, 1.20, 1.22, 1.24, 1.26 \}$.
 Statistical errors for the CDF method were calculated with 80 jackknife bins.}
\label{fig:3}
\end{figure}

\subsection{Ubiquitin}

    Ubiquitin is a 76 residue small protein (PDB code 1UBQ) for which evidences suggest a two-state folding mechanism.
    It had received considerable attention for what concerns solvent response and temperature dependence in the secondary 
structure formation
\cite{ubqJacob,chung-ubiquitin,ghosh-ubiquitin} and stretching experiments 
\cite{jernigan-ubiquitin,das-ubiquitin,imparato-ubiquitin}.

    To study the performance of the numerical approaches in evaluating $\beta(E)$ through the folding-unfolding
transition, we perform simulations with a coarse-grained version for Ubiquitin but with a rather detailed potential.
    We use the structure-based model implemented in SMOG@ctbp web server \cite{smog} to perform MD simulations
with GROMACS package.
    The analyses were made from $10^6$ measurements during 250~ns of simulation after a period of thermalization.
    Measurements were obtained from 7 independent simulations at temperatures $T_{\alpha}=T_1+(\alpha-1)\,\Delta T$ 
with $T_{1}=1.14$ and $\Delta T=0.02$.

    First, we evaluate the performance of the approaches considering the data obtained from a single MD trajectory
produced at $T_4=1.20$, which is close to the transition temperature $T_c=1.219(3)$ evaluated with statistics
obtained with 7 independent MD simulations. 
    Figure \ref{fig:1} displays the estimates for the caloric curve using the linear regression to evaluate
derivatives of  ${\rm ln}H(E_m)$ from data obtained at $T=1.20$.  
    This figure shows how noisy the caloric curve can be as a function of the energy binsize $\varepsilon = 0.3, 0.8$ and 1.3,
keeping  the number of points $k=15$ fixed.
    Figure \ref{fig:2} compares what seems a good energy discretisation with the CDF method.    	   
    Error bars based on 20 jackknife bins are presented only for the CDF method. 
    Now, Fig. \ref{fig:3} compares both methods when we consider the statistics obtained with the entire set of 
temperatures $T_\alpha$, $(\alpha=1,2, \cdots, 7)$.
    To present error bars for the CDF method, we have increased the number of jackknife bins to 80 because
this larger statistics.
    Nice agreement between both methods are obtained if we choose $\varepsilon$ conveniently.

\begin{figure}[ht]
\begin{center}
\includegraphics[width=0.48\textwidth]{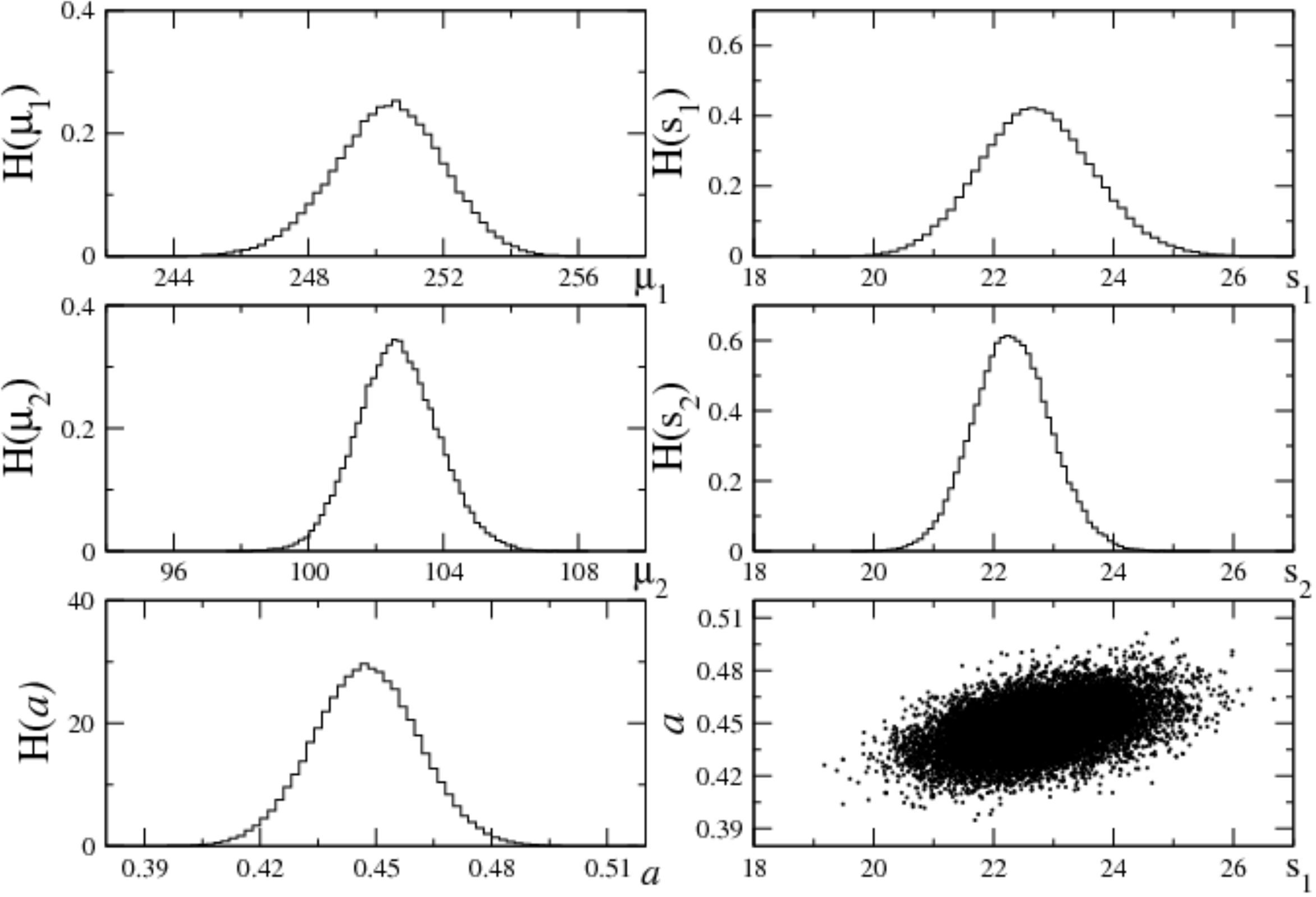}
\end{center}
\caption{Marginalized PDFs of the parameters for model I, and scatter plot for joint probability of $a$ and $s_1$
         with data obtained for Ubiquitin at $T=1.20$.}
\label{fig:6}
\end{figure}

    Figure \ref{fig:4} displays what we call input experimental points to perform the Bayesian analysis.
    This {\it experimental} data was obtained from MD simulations at $T=1.20$.
    The dashed line corresponds to the cumulative distribution obtained with model I.
    The parameters of this model that fit the input data are displayed in Table 1. 
    We include values calculated from means of the marginalized distributions, and global modes that presented
the smallest $\chi^2$/d.o.f of the model.
    To verify how the proposed model describes the energy distribution when compared with the CDF method
and naive histogramming, we show in Fig. \ref{fig:5} the results obtained from these methods for the statistics 
collected at $T=1.20$.
    We realize that model I does not recover properly the region between the peaks of the energy
distributions $H(E)$ obtained with previous numerical methods.
    The inset in this figure compares the calculations of $dH(E)/dE$ following from the CDF and Bayesian approaches.
    In Fig. \ref{fig:6} we display the marginalized distributions of $\vec{\lambda}$, including the
correlation between the parameters $s_1$ and $a$ to illustrate their interdependence.
    We have always used flat {\it priors} over appropriated ranges to obtain the parameter distributions.

\subsection{Fibonacci sequences}

\begin{figure}[ht]
\begin{center}
\begin{minipage}[h]{0.45\textwidth}
\includegraphics[width=0.95\textwidth]{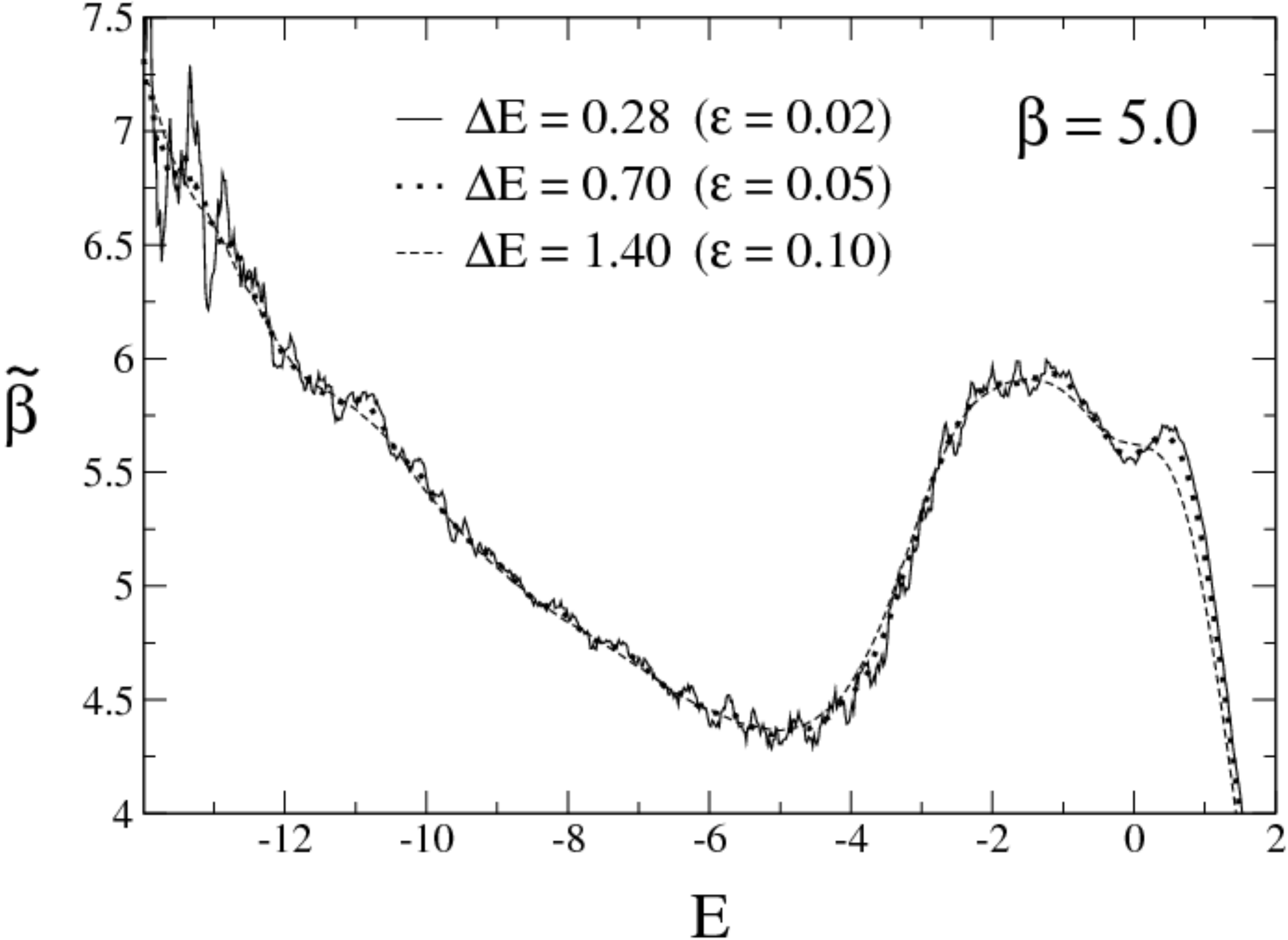}
\caption{Estimates ${\tilde \beta}(E)$ for the aggregation of two Fibonacci sequences 
for different parameters $\varepsilon$ and $k=15$ points with
data selected for $\beta=5.0$ among time series produced with REM.}
\label{fig:7}
\end{minipage}%

\vspace{0.8cm}

\begin{minipage}[h]{0.45\textwidth}
\includegraphics[width=0.95\textwidth]{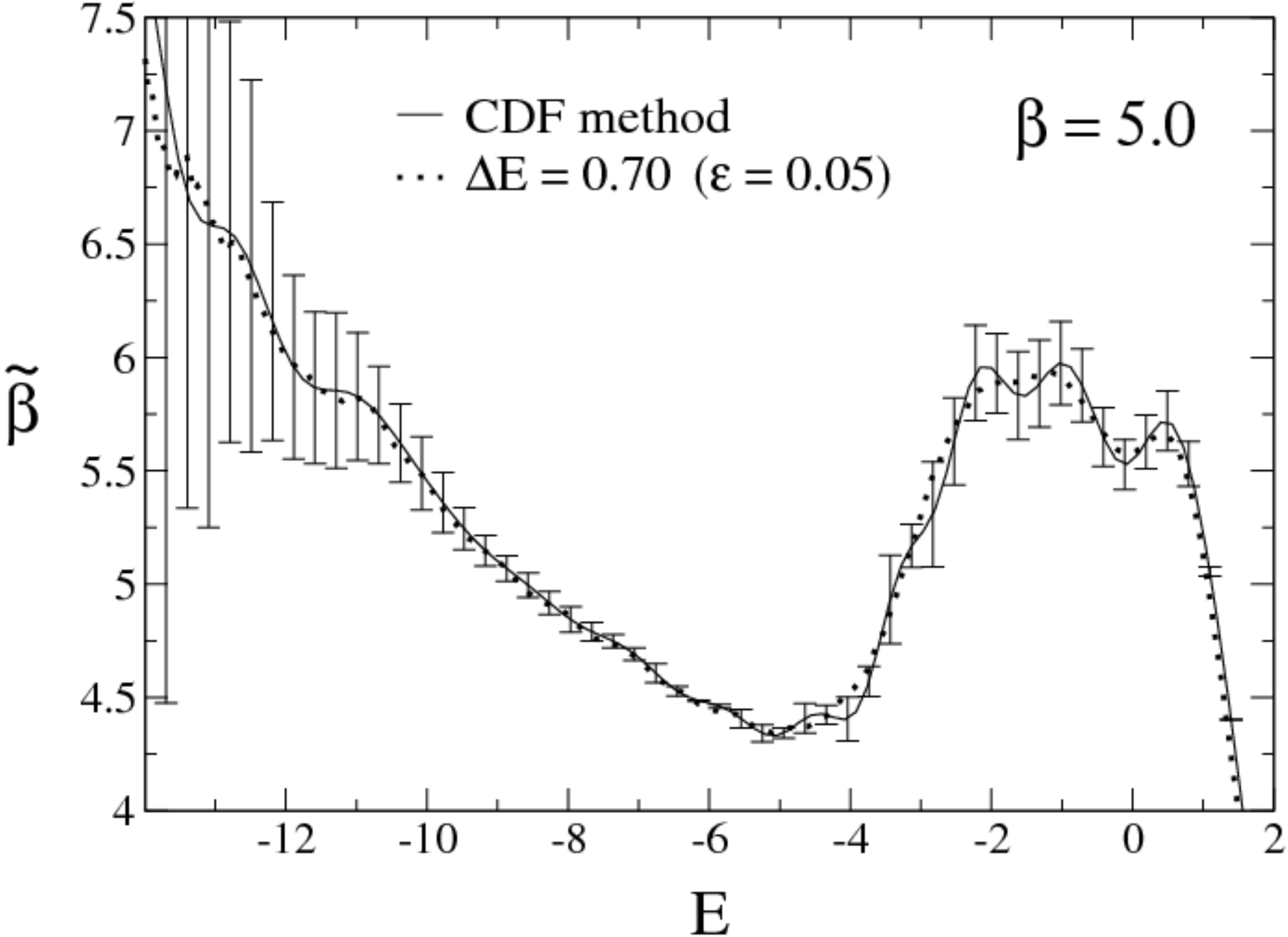}
\caption{Comparison between estimates ${\tilde \beta}(E)$ for the aggregation of two Fibonacci 
sequences with the linear regression 
($\varepsilon= 0.05$) and CDF method for data obtained at $\beta=5.0$. 
Statistical errors for the CDF method were calculated 
with 20 jackknife bins.}
\label{fig:8}
\end{minipage}%
\end{center}
\end{figure}

    Figure \ref{fig:7} shows the caloric curve for the aggregation of two chains of monomers 
described by the AB model, a coarse-grained off-lattice protein model that replaces the all-atom potentials by simpler 
physical interactions.
    This model reduces the protein to a chain of monomers of two types: the hydrophobic ($A$) and the polar or 
hydrophilic ($B$) types, located at the $C_{\alpha}$ atoms.
    The AB model has also been used in studies to explore aggregation phenomenon \cite{jankePRL2006,jankeJCP2008}, 
where more than one chain is included in the system, and in studies of microcanonical thermostatistics of 
heteropolymers \cite{frigoriJCP2013}.

    Here, we consider a simple system which consists of two heteropolymers defined by Fibonacci sequences with $13$ 
monomers, {\it i.e.} {\ttfamily ABBABBABABBAB}.
    The statistics for this system amounts to $10^7$ sweeps per replica produced with REM.
    Attempts to exchange the replicas occur every $n_s=10^4$ sweeps, using a scheme that alternates attempts  
between even replicas and their neighbors, and odd replicas and their neighbors.
    Although one has several manners to define the temperature set \cite{fiore}, here we determine it using an arithmetic 
progression $\beta_{\alpha}=\beta_{1} + (\alpha-1)\,\Delta \beta$.
    We consider $N_{\rm rep}=7$ replicas choosing $\beta_{1}=4.2$, and $\Delta \beta=0.4$.

\begin{figure}[ht]
\begin{center}
\includegraphics[width=0.45\textwidth]{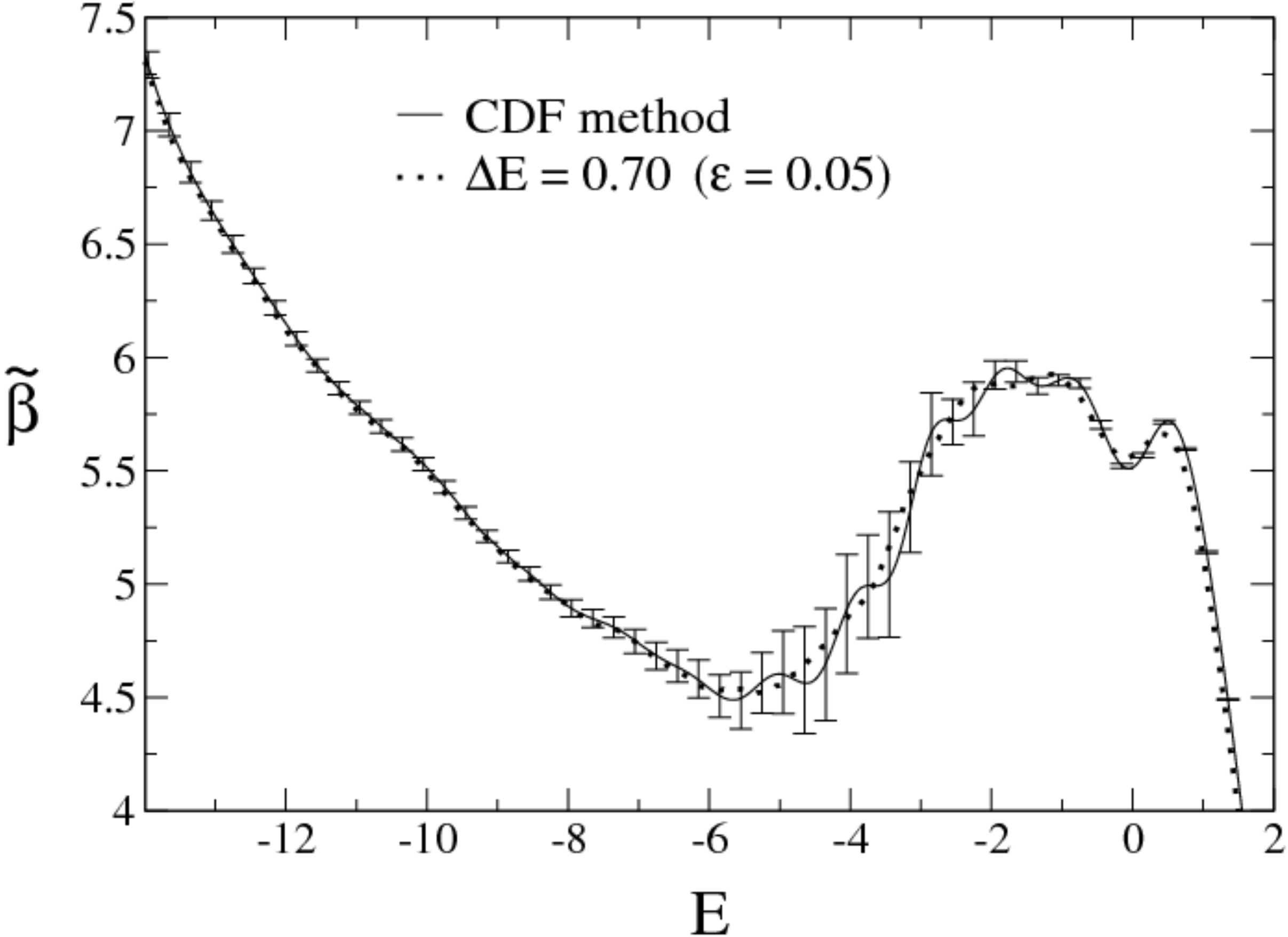}
\end{center}
\caption{Comparison between estimates ${\tilde \beta}(E)$
for the aggregation of two Fibonacci sequences with the linear regression 
($\varepsilon= 0.05$) and CDF method for energy time series 
obtained from REM at $\beta = \{ 4.2, 4.6, 5.0, 5.4, 5.8, 6.2, 6.6 \}$.}
\label{fig:9}
\end{figure}

    Figure \ref{fig:7} depicts the caloric curves ${\tilde \beta}(E)$ obtained with energy binsizes  
$\varepsilon=0.02, 0.05$, and $0.10$.
    The value $k=15$ was used to calculate the derivatives of ${\rm ln} H(E_m)$.
    Different values of $\varepsilon$ allows one to verify the systematic errors associated with
$\varepsilon$ and the use of linear regression.
    For a fixed $k$,  small values of $\varepsilon$ (or $\Delta E$) tend to reproduce the 
statistical fluctuations of the data.
    On the other hand, a value as large as $\varepsilon=0.10$ eliminates the physical information
associated with the small S-loop at $E \sim 0.2$.
    For comparison, we include in Fig. \ref{fig:8} the ${\tilde \beta}(E)$ estimates from the CDF method.
    These comparisons are based on a single time series selected from data produced with REM at 
the inverse temperature $\beta = 5.0$.
    It is important to reliably calculate the caloric curve when assessing the canonical critical temperature and 
the latent heat of the transition.
    We can figure out how important the fluctuations observed in these curves ${\tilde \beta}(E)$ are by 
calculating the statistical errors associated with the CDF method.
    Figure \ref{fig:8} displays the statistical error bars calculated with 20 jackknife bins
for the dataset with $10^7$ measurements obtained with REM at $\beta=5.0$.
    Both methods present comparable statistical error bars. 
    This figure shows that the small S-loop at $E \sim 0.2$ does not result from the statistical fluctuations in
our dataset.
    Therefore, any smooth estimate of this curve at $E \sim 0.2$, obtained for example with $\Delta E = 1.40$ 
($\varepsilon = 0.10$), would hide physical information.

\begin{figure}[!t]
\begin{center}
 \begin{minipage}[h]{0.45\textwidth}
\includegraphics[width=0.95\textwidth]{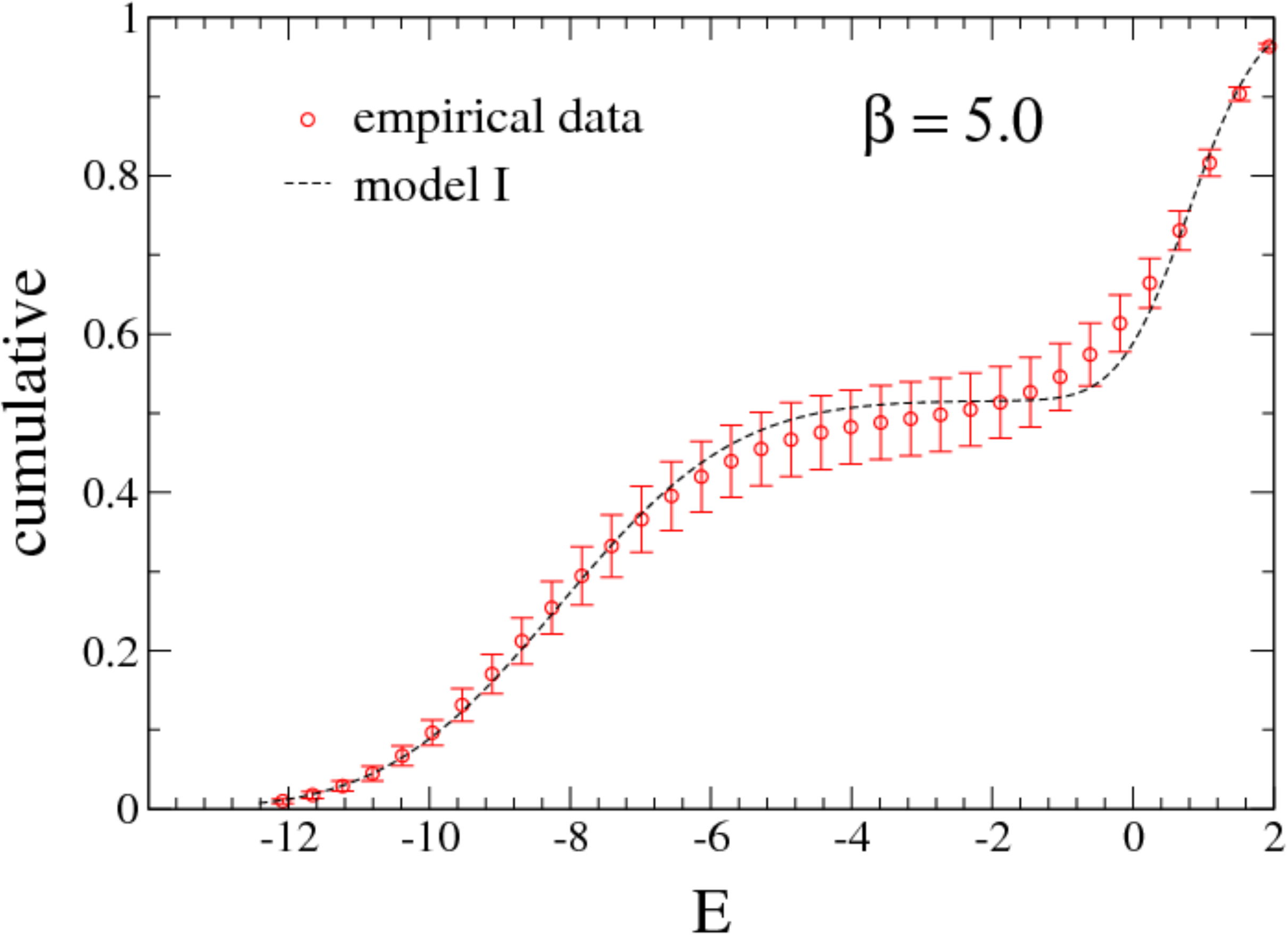}
\caption{Cumulative distribution data ($\circ$) for the aggregation of two Fibonacci 
sequences with error bars calculated with 20 jackknife bins from
REM simulation for the energy time series obtained at $\beta=5.0$. 
Dotted line corresponds to the Bayesian estimates of the cumulative distribution
assuming model I.}
\label{fig:10}
\end{minipage}%

\vspace{0.8cm}

\begin{minipage}[h]{0.45\textwidth}
\includegraphics[width=0.95\textwidth]{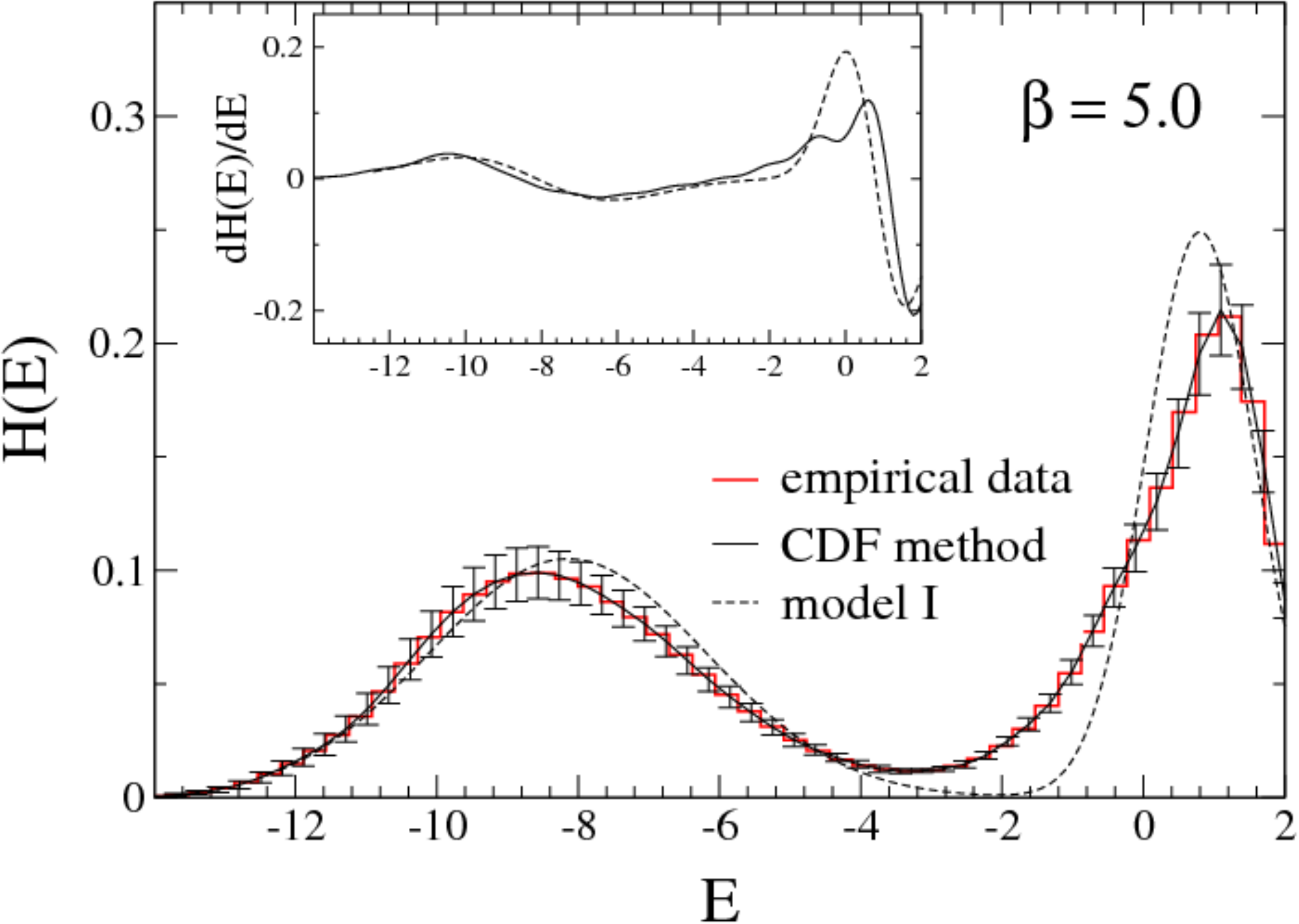}
\caption{Comparison of energy histograms $H(E)$ for the aggregation of two Fibonacci 
sequences. Error bars are calculated with 20 jackknife bins are shown for the CDF method. 
Inset: comparative behaviour of $dH(E)/dE$ as predicted with the CDF method and Bayesian analysis
assuming model I.}
\label{fig:11}
\end{minipage}%
\end{center}
\end{figure}

    In Fig. \ref{fig:9} we include data produced with $N_{\rm rep} =7$ replicas together.
    Comparison of the results presented in this figure with those obtained by multicanonical simulations in 
Refs. \cite{jankePRL2006,jankeJCP2008} reveals that they are in excellent agreement:
even the small S-loop near $E\sim 0.2$ appears in our case.
    Computation of the inverse of aggregation temperature $\beta_{\rm agg} = 5.212(59)$ also resulted in a 
quantitative agreement.    
    Larger error bars appear around $E =-4$, probably because the lack of temperatures in the 
set used to exchange the replicas.
    The averaged estimates from both derivative calculation methods agreed
for a particular choice of $\varepsilon=0.05$ and $k=15$ points in the linear regression method.
 		
    Now, Fig. \ref{fig:10} shows the {\it experimental} data we have used to perform the Bayesian analysis.
    Again, we initially keep the analysis restricted to a single energy time series, in this case produced at $\beta=5.0$.
    We have included the Bayesian estimates (dashed line) of the cumulative function assuming the model I.
    The parameters for this model are presented in Table I.
    To verify how well this model reproduces $P(E)$, we include in Fig. \ref{fig:11} the energy distributions
obtained with a naive histogramming procedure, CDF method and Bayesian analysis.
    Clearly, Bayesian analysis does not reproduce the expected behaviour between both peaks.
     This leads to a biased estimation of $dH(E)/dE$ as can be seem in the inset of this figure.
     Moreover, the approach with model I does not reproduce the small S-loop observed at $E \sim 0.2$
(data not shown), in contrast to the estimates with a convenient choice of $\varepsilon$ or the CDF method.
     As a consequence, we again do not follow further analysis with the Bayesian approach including all 
statistics from the 7 replicas.

\section{Summary and conclusions}

       We presented two alternative approaches for the estimation of the energy probability distributions 
 avoiding the need of energy binning in order to obtain the microcanonical thermostatistics analysis from ST-WHAM.
       Numerical comparisons between the approaches were presented and we showed the statistical and systematic errors 
that can arise when one evaluates the microcanonical inverse temperature for two continuous energy models that exhibits 
first-order like phase transitions.
	Our results indicate that the CDF method yields reliable estimates for both $H(E)$ and $\beta(E)$ when compared with the 
linear regression method, and far more successfully than the Bayesian approach with model I.
	Unlike the linear regression method, the CDF method avoids the undesirable task of choosing the energy binsize 
$\varepsilon$ for a careful evaluation of the microcanonical 
temperature, as highlighted in the analysis of the aggregation transition of the two Fibonacci sequences.
	In this case, the caloric curve presents a quite unusual behaviour with two S-loops, indicating two transitions.
	In particular, we showed that the use of large values for $\varepsilon$ in linear regression method would hide physical 
information about the small S-loop at $E \sim 0.2$.
	On the other hand, the small S-loop could not be detected with the Bayesian analysis (data not shown) as a 
consequence of the poor evaluation of $H(E)$ and its derivative (Fig. \ref{fig:11}).
    The reason is because the model I consists of only 5 parameters.
    This is a low number compared to the CDF method, which allows a variable number of parameters 
({\it i.e.} Fourier coefficients) defined depending on the demand of the ECDF.
    For instance, the CDF method needs 13 Fourier coefficients to obtain probability densities for the Ubiquitin data 
and this number goes up to 74 for Fibonacci sequences. 
    Furthermore, model I was constructed on the hypothesis that the coexistence of two thermodynamic bulk phases can be well 
described by two gaussian distributions.
    Actually, this is an approximation because the energies in between the two peaks, describing
mixed phase configurations, are not properly take into account in the two gaussian model.

\section*{Acknowledgments}
 We thank Bernd Berg for carefully reading our manuscript. 
 This work has been supported by the Brazilian agencies FAPESP, CNPq and RUSP (University of S\~ao Paulo).

\section{Appendix}

 Derivative of the probability density:

{\small 
\begin{verbatim}
 FUNCTION DERPD(X)  
 include '../../Libs/Fortran/implicit.sta' 
 include '../../Libs/Fortran/constants.par' 
 PARAMETER(NMAX=100)       
 COMMON /CDFProb/ XMIN,XRANGE,DN(NMAX),M  
!M    number of Fourier coefficients
 DERPD=ZERO                              
 DO J=1,M                      
   AUX=J*PI/XRANGE                
   AUX=AUX*AUX              
   DERPD=DERPD-DN(J)*AUX*SIN(J*PI/XRANGE*(X-XMIN))  
 ENDDO                  
 RETURN               
 END                    
\end{verbatim}
}



\begin{table}[!b]
\begin{minipage}[h]{1.0\textwidth}
\begin{center}
\small\addtolength{\tabcolsep}{-2pt}
\begin{tabular}{cclllllc}
\hline \\
[-0.35cm]
\hline \\
[-0.35cm]
Examples  &             & ~~~~$\mu_1$    &  ~~~~$s_1$     & ~~~~$\mu_2$   & ~~~~$s_2$     & ~~~~$a$        & ~~~~$\chi^2$/d.o.f.   \\
\hline \\
[-0.15cm]
Ubiquitin & ~~~~mean        & ~~~~250.36(37) & ~~~~22.73(22)  & ~~~~102.64(27) & ~~~~22.31(15) & ~~~~0.4474(31) &  ~~~~0.131           \\
          & ~~~~global mode & ~~~~250.51     & ~~~~22.64      & ~~~~102.55     & ~~~~22.27     & ~~~~0.4474     &  ~~~~0.130           \\
Fibonacci & ~~~~mean        & ~~~~0.797(15)  & ~~~~0.7786(79) & ~~~~-8.128(37)  & ~~~~1.968(25) & ~~~~0.4838(35) &  ~~~~0.350            \\
sequences & ~~~~global mode & ~~~~0.803      & ~~~~0.7754     & ~~~~-8.147      & ~~~~1.958     & ~~~~0.4841     &  ~~~~0.349            \\  
 \hline \\
[-0.35cm]
\hline
\end{tabular}
\caption{
Parameters for the proposed model in Eq. (\ref{eq:fe}) from averages and the $\chi^2$/d.o.f of the model.
}
\label{tab:bayes}
\end{center}
\end{minipage}
\end{table}

\end{document}